\documentclass[twocolumn,useAMS,usenatbib]{mn2e}
\usepackage{amsmath,latexsym,amssymb}
\usepackage{epsfig,graphicx}
\def\reff@jnl#1{{\rm#1\/}}

\def\aj{\reff@jnl{AJ}}                  
\def\araa{\reff@jnl{ARA\&A}}            
\def\apj{\reff@jnl{ApJ}}                        
\def\apjl{\reff@jnl{ApJ}}               
\def\apjs{\reff@jnl{ApJS}}              
\def\apss{\reff@jnl{Ap\&SS}}            
\def\aap{\reff@jnl{A\&A}}               
\def\aapr{\reff@jnl{A\&A~Rev.}}         
\def\aaps{\reff@jnl{A\&AS}}             
\def\baas{\reff@jnl{BAAS}}              
\def\jrasc{\reff@jnl{JRASC}}            
\def\memras{\reff@jnl{MmRAS}}           
\def\mnras{\reff@jnl{MNRAS}}            
\def\physrep{\reff@jnl{Phys.Rep.}}
\def\pra{\reff@jnl{Phys.Rev.A}}         
\def\prb{\reff@jnl{Phys.Rev.B}}         
\def\prc{\reff@jnl{Phys.Rev.C}}         
\def\prd{\reff@jnl{Phys.Rev.D}}         
\def\prl{\reff@jnl{Phys.Rev.Lett}}      
\def\pasp{\reff@jnl{PASP}}              
\def\pasj{\reff@jnl{PASJ}}              
\def\skytel{\reff@jnl{S\&T}}            
\def\solphys{\reff@jnl{Solar~Phys.}}    
\def\sovast{\reff@jnl{Soviet~Ast.}}     
\def\ssr{\reff@jnl{Space~Sci.Rev.}}     
\def\nat{\reff@jnl{Nature}}             

\def\sun{\hbox{$\odot$}}

\newcommand{\hmpc}{\ensuremath{h^{-1}\mathrm{Mpc}}}

\newcommand{\beq}{\begin{equation}}
\newcommand{\eeq}{\end{equation}}
\newcommand{\beqa}{\begin{eqnarray}}
\newcommand{\eeqa}{\end{eqnarray}}

\title[Cluster intrinsic alignments]{Detection of intrinsic cluster
  alignments to $100h^{-1}$Mpc in the SDSS}

\author[Smargon et al.]
{A. Smargon$^1$, 
R. Mandelbaum$^{1,2}$\thanks{\tt rmandelb@astro.princeton.edu},
N. Bahcall$^1$,
M. Niederste-Ostholt$^3$
\\$^1$Department of Astrophysical Sciences, Princeton University,
Peyton Hall, Princeton, NJ 08544, USA
\\$^2$Department of Physics, Carnegie Mellon University, Pittsburgh,
PA 15213, USA 
\\$^3$Institute of Astronomy, Madingley Rd, Cambridge, CB3 0HA
}

\date{\today}

\begin{document}

\bibliographystyle{mn2e}
\maketitle

\begin{abstract}
  We measure the large-scale intrinsic alignments of galaxy clusters
  in the Sloan Digital Sky Survey (SDSS) using subsets of two cluster
  catalogues: 6625 clusters with $0.1<z<0.3$ from the 
  maxBCG cluster catalogue \citep[][7500 deg$^2$]{2007ApJ...660..221K}, and 8081
  clusters with $0.08<z<0.44$ from the Adaptive Matched
  Filter catalogue \citep[][6500 deg$^2$]{2008ApJ...676..868D}.  We search for two
  types of cluster alignments using pairs of clusters: the alignment
  between the projected major axes of the clusters (`correlation'
  alignment), and the alignment between one cluster major axis and the
  line connecting it to the other cluster in the pair (`pointing'
  alignment).  In each case, we use the cluster member galaxy
  distribution as a tracer of the cluster shape.  All measurements are
  carried out with each catalogue separately, to check for dependence
  on cluster selection procedure.  We find a strong detection of the
  pointing alignment on scales up to $100h^{-1}$Mpc, at the $6$ or
  $10\sigma$ level depending on the cluster selection algorithm used.
  The correlation alignment is only marginally detected up to $\sim
  20h^{-1}$Mpc, at the $2$ or $2.5\sigma$ level.  These results
  support our current theoretical understanding of galaxy cluster
  intrinsic alignments in the $\Lambda$CDM paradigm, although further
  work will be needed to understand the impact of cluster selection
  effects and observational measurement errors on the amplitude of the
  detection.
\end{abstract}


\section{Introduction}

The $\Lambda$CDM cosmological paradigm features a cosmic web
containing galaxies, filaments, galaxy clusters, and larger
superclusters.  These structures are hosted by dark matter halos which
are predicted to have shapes that are aligned with each other due to
tidal forces and coherent matter infall  along filaments
\citep{2000ApJ...545..561C,2000MNRAS.319..649H,2001MNRAS.320L...7C,2001ApJ...559..552C,2002MNRAS.335L..89J,2005ApJ...618....1H}.
These alignments are manifested in various ways: for example, as alignments of
galaxy or galaxy cluster shapes towards overdensities; alignments of
pairs of galaxy or cluster 
shapes with each other; and alignments of galaxy cluster shapes with 
the shape of the Brightest Cluster Galaxy (BCG).

Recent intrinsic alignments work has focused on 
galaxies, in
part because of the pernicious effects of such alignments on weak
lensing measurements (e.g.,
\citealt{2000ApJ...545..561C,2004PhRvD..70f3526H}). This work 
suggests that the large-scale ($10$--$100h^{-1}$ Mpc) intrinsic
alignments of galaxies are a complex function of 
luminosity, colour and/or morphological type (e.g., \citealt{2006MNRAS.367..611M,2007MNRAS.381.1197H}).  The non-detection of
intrinsic alignments for lower luminosity galaxies ($\lesssim L*$) may result from 
misalignment between the galaxy light distribution and the
underlying dark matter halo shape
\citep{2006MNRAS.371..750H,2011arXiv1108.3717B}. 

If we want to measure intrinsic alignments with fewer systematic
errors, we might instead consider the projected (2d) intrinsic alignments of galaxy
clusters.  The cluster member galaxies trace the cluster shape
sufficiently well that they should be useful for testing the
$\Lambda$CDM predictions for intrinsic alignments of cluster-scale
dark matter halos (as suggested by a successful stacked weak lensing measurement of cluster
ellipticity in the SDSS, \citealt{2009ApJ...695.1446E}, which is only
feasible if the cluster shape traced by the member galaxies strongly
correlates with the shape of the underlying dark matter halo). There are several sets of theoretical predictions
\citep{1997ApJ...479..632S,2000MNRAS.319..614O,2002A&A...395....1F,2005ApJ...618....1H}
that qualitatively agree that cluster intrinsic alignments should extend to
$\sim 100h^{-1}$Mpc scales for a $\Lambda$CDM cosmology (though minor disagreements arise due to
different $N$-body simulation volumes, resolutions, and cosmologies).
Additional theoretical work has quantified the environment-dependence
of such alignments; e.g., \cite{2006MNRAS.370.1422A} found that the
shapes of nearby clusters are aligned if the clusters are connected by
a filament.  Thus, matter infall along filaments is an important
factor in galaxy cluster intrinsic alignments (see also \citealt{2005ApJ...618....1H}).

There were several early attempts to measure intrinsic
alignments of galaxy cluster shapes up to tens of Mpc scales \citep{1982A&A...107..338B,1985AJ.....90..582S,1987MNRAS.228..941F,1988AJ.....95..996L,1989ApJ...338..711U,1989ApJ...344..535W,1994ApJS...95..401P}.
Typically, they used small, inhomogeneous cluster samples; results were conflicting and the reported
detections were typically low in significance, and limited to $\lesssim 30
h^{-1}$Mpc.  

Recent work with has 
demonstrated several smaller-scale cluster alignments.  For
  example, alignments of groups within the local supercluster have
  been demonstrated up to 20 Mpc scales \citep{2010ApJ...708..920G}.  
Using larger samples, the alignment of the shape of the BCG or the X-ray isophotes with
that of the host cluster has been robustly demonstrated
\citep{1999ApJ...519...22F,2002ASPC..268..395K,2008MNRAS.390.1562H,2010MNRAS.405.2023N,2011ApJ...740...39H}.
Moreover, correlations have been detected up to $30h^{-1}$Mpc between
the cluster X-ray isophotes and the position of the
nearest neighbour cluster, or the density field traced by spectroscopic
galaxy positions 
\citep{2000ApJ...544..104C,2002ApJ...565..849C,2009ApJ...703..951W,2011MNRAS.414.2029P}.
 Given these promising results
with larger, more homogeneous cluster samples, we attempt to detect, for
the first time, the intrinsic alignments of galaxy clusters on the largest
scales for which there is a theoretical prediction ($100h^{-1}$Mpc),
using  SDSS data.

\section{Data}

The SDSS \citep{2000AJ....120.1579Y} imaged roughly $\pi$ steradians
of the sky by drift-scanning the sky in photometric conditions
\citep{2001AJ....122.2129H, 2004AN....325..583I} in five bands
($ugriz$; \citealt{1996AJ....111.1748F, 2002AJ....123.2121S}) using a
specially-designed wide-field camera \citep{1998AJ....116.3040G}. The data were processed by automated pipelines that
detect and measure photometric properties of objects, and
astrometrically calibrate the data \citep{2001ASPC..238..269L,
  2003AJ....125.1559P,2006AN....327..821T}.  This paper relies on SDSS
Data Release 6 \citep[DR6,
][]{2008ApJS..175..297A}.

We use two pre-existing catalogues of galaxy clusters based on
photometric data.  The first is
the maxBCG catalogue \citep{2007ApJ...660..221K,2007ApJ...660..239K},
which includes 12~766 clusters in 7500~deg$^2$, with $0.1<z<0.3$,
detected by searching for an overdensity of red galaxies.  This
catalogue includes clusters with $\ge 10$ red sequence member galaxies
with $L\ge 0.4L_*$ within the cluster radius $R_{200}$.   
The second catalogue \citep{2008ApJ...676..868D} results from using
the 
Adaptive Matched Filter (AMF) algorithm to identify galaxy
overdensities (not necessarily red) over a larger
redshift range, yielding 36~785 clusters in 6500~deg$^2$, each with $\ge 20$ member galaxies
with $L\ge 0.4L_*$. 
Both
catalogues have photometric redshifts with typical uncertainty $\Delta
z_\mathrm{phot}\sim 0.015$, which complicates the identification of
nearby cluster pairs.  Comparison of results for the two catalogues may indicate
how differences in the cluster selection procedure affect our ability
to measure intrinsic alignments.

The analysis requires measurements of the cluster shapes.  We use
the shapes measured by \cite{2010MNRAS.405.2023N}, who identified red
cluster member galaxies within $0.5$~Mpc of the geometric cluster
centre, and measured cluster shapes using the radius-weighted second
moments of the galaxy distribution.  To obtain a sample with reliable
shapes, \cite{2010MNRAS.405.2023N} recommend a series of cuts on the
clusters as shown in Table~\ref{T:cuts}, resulting in final catalogues of
$8081$ and $6625$ clusters with shapes for AMF and maxBCG, 
respectively.  The cut to the redshift range $0.08<z<0.44$, which is
necessary for robust identification of the red sequence,  only affects
the AMF catalogue, since the maxBCG catalogue is limited to a narrower
redshift range already.  One cut, requiring that the cluster ellipticity be
inconsistent with zero at $>1\sigma$, may introduce a potential bias
on the systems that are selected.  The original reason for
  this cut is that, for a given number of member galaxies, as the
  ellipticity appears rounder, the position angle that is estimated
  becomes noisier.  Extremely noisy position angles will wash out an
  alignment signal that might actually be present if we were able to
  measure the cluster shape to arbitrary precision.

While we require that $\ge 5$
member galaxies are used to determine the shapes, only 5--6
were used for the majority of the clusters\footnote{This $\ge 5$ cut
  used for the determination of shapes may seem inconsistent with the
  quoted richness thresholds for the AMF and maxBCG cluster catalogues
of $20$ and $10$ cluster members.  In fact, the cut used for shape
determination is more stringent, both because it requires that the
member galaxies be red (unlike the AMF richness estimator) and, more
importantly, that they lie within $0.5$Mpc of the cluster centre,
rather than within the virial radius which may be several times
larger.}.  The small number of galaxies used to determine shapes may introduce some
random noise into the position angles (as in figure 10 of \citealt{2010MNRAS.405.2023N}), which will tend to dilute
intrinsic alignment signals.

\begin{table*}
\begin{tabular}{lcc}
\hline
 & $\!\!$AMF$\!\!$ & $\!\!$maxBCG \\
 & $\!\!\!$clusters$\!\!\!$ & clusters \\
\hline
Input catalogue & 36~785$\!\!$ & $\!\!$12~766 \\
$0.08<z<0.44$ (AMF), $0.1<z<0.3$ (maxBCG) & 23~106$\!\!$ & $\!\!$12~766 \\
Geometric centre $<0.5$ Mpc from input centre$\!\!\!\!\!$ & 21~711$\!\!$ & $\!\!$12~202 \\
BCG within 0.5~Mpc of geometric centre & 14~053$\!\!$ & $\!\!$10~754 \\
Inconsistent with being round at $>1\sigma$ & 9115 & 7071 \\
$\ge 5$ member galaxies used to define shape & 8081 & 6625 \\
\hline
\end{tabular}
\caption{Cuts imposed on clusters in both cluster catalogues, in order
to obtain reliable position angle measurements (from  \citealt{2010MNRAS.405.2023N}).\label{T:cuts}}
\end{table*}

\section{Intrinsic alignments estimator}\label{S:estimator}

To begin, we isolate cluster pairs that may be associated (in 3d).
For each possible cluster pair, we compute its angular separation on
the sky, $\Delta\phi$, and require it to be consistent with a comoving
separation of $R\le 100h^{-1}$Mpc at the mean photometric redshift of the pair
(using a fiducial flat $\Lambda$CDM cosmology with $\Omega_m=0.27$,
and expressing our results with $h=1$).  Along the line-of-sight, we
require $\Delta z = |z_1-z_2| \le 0.015$, a redshift separation
corresponding to the $1\sigma$ photometric redshift error (the impact
of this selection will be discussed in Sec.~\ref{S:theory}).  There are
a total of 103~175 and 58~899 cluster pairs for the maxBCG and AMF
catalogues, respectively.  Each cluster can in principle be included
in several cluster pairings.

We divide the cluster pairs into 9 logarithmically spaced bins in $R$, from
$1$ to $100h^{-1}$Mpc.  For each pair, we compute two statistics defined in \cite{2005ApJ...618....1H}:  the `correlation angle'
$\theta_c$ is the angle between the projected major axes of the two clusters,
and the `pointing angle' $\theta_p$ is the angle on the sky between
the projected cluster
major axis and the line connecting one cluster to the other
cluster in the pair.  Thus, $\theta_c$ indicates the alignment of
the cluster shapes with each other, and $\theta_p$ indicates whether
clusters tend to point towards other clusters.  In each $R$ bin, we compute
the mean $\langle \cos^2{\theta_c}\rangle$ and likewise for
$\theta_p$; this statistic would be $0.5$ for a
purely random distribution of cluster shapes.  

We estimate errorbars by assigning each cluster a random orientation
angle.  We then redo the measurement procedure with the random angles,
in which case the ideal signal is a known value,
$\langle\cos^2{\theta_{c,p}}\rangle = 0.5$. Deviations from that value
can be used to quantify the noise in
$\langle\cos^2{\theta_{c,p}}\rangle$ in the real measurement due to
the finite number of cluster pairs used for the analysis.  The
errorbars shown on the plot are the standard deviation of the mean
$\cos^2{\theta_{c,p}}$.

\section{Observational results}\label{S:results}

\begin{figure}
\includegraphics[width=\columnwidth]{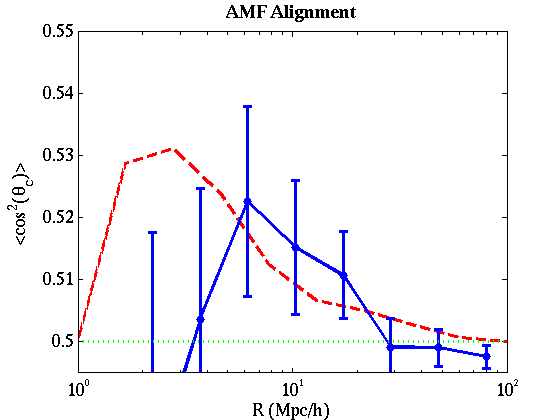} 
\includegraphics[width=\columnwidth]{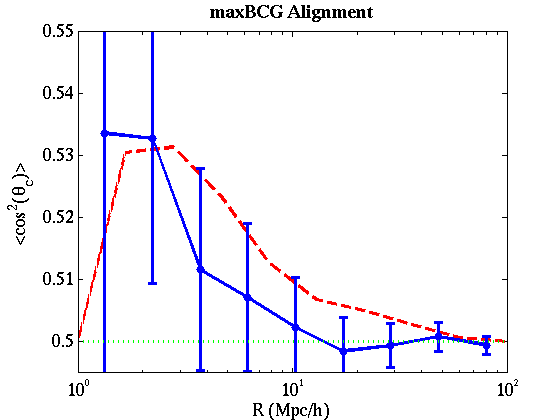} 
\caption{The cluster correlation angle alignment 
  $\langle\cos^2{\theta_c}\rangle$, as a function of comoving pair separation
  $R$, for AMF (top) and maxBCG (bottom)
  catalogues.   The blue points with errorbars are the observational
  results (Sec.~\ref{S:results}); the red dashed lines are theoretical
predictions corrected for photometric redshift uncertainties
(Sec.~\ref{S:theory}); and the green dotted horizontal lines indicate
purely random cluster orientations.  The errors between the different
bins in $R$ are independent.}
\label{fig:corr_results}
\end{figure}

In this section, we present the results of the measurement described in Sec.~\ref{S:estimator}.
Fig.~\ref{fig:corr_results} shows the correlation angle alignment for
both cluster catalogues.  As shown, the correlation angle
alignment is detected below $20h^{-1}$Mpc, but with large error bars, and represents a $2.5\sigma$ (AMF) and
$2\sigma$ (maxBCG)
detection summed over these radial bins.  The suppressed signal for $R<3h^{-1}$Mpc in the AMF
catalogue may result from the difficulty in identifying 
cluster pairs with small separation and measuring their shapes, given that this separation
is comparable to the typical cluster size.  It is unclear why the
maxBCG algorithm seems to do slightly better in this respect.

\begin{figure}
\includegraphics[width=\columnwidth]{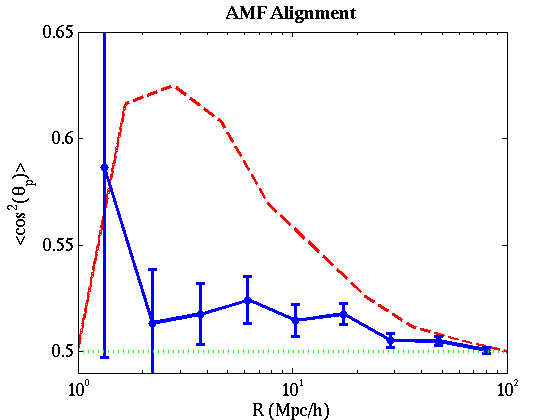} 
\includegraphics[width=\columnwidth]{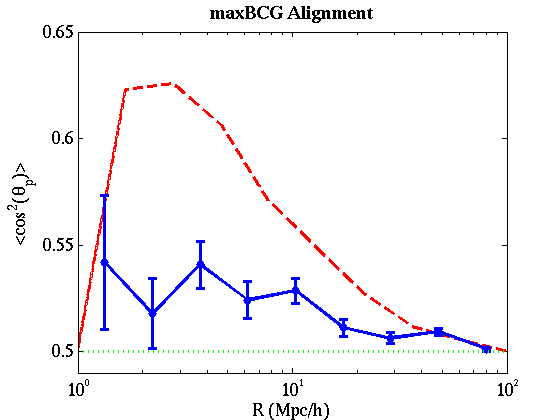} 
\caption{The cluster pointing angle alignment 
  $\langle\cos^2{\theta_p}\rangle$, as a function of comoving
  pair separation $R$, for AMF (top) and maxBCG (bottom)
  catalogues. The line format is the same as in Fig.~\ref{fig:corr_results}.}
\label{fig:point_results}
\end{figure}
Fig.~\ref{fig:point_results} shows the measured pointing angle
alignments, for which there is clearly a signal on all scales,
indicating that clusters tend to point towards other clusters out to
$100h^{-1}$Mpc.  This signal is detected at $6\sigma$ (AMF) and
$10\sigma$ (maxBCG) averaged over $1<R<100h^{-1}$Mpc
scales, with the detection significance remaining strong beyond
50\hmpc\ but dropping to $1\sigma$ in the outermost bin.  

The errors are smaller for the pointing angle measurement than for the
correlation angle (Fig.~\ref{fig:corr_results}) by roughly a factor of
two, for two reasons.  First, the correlation angle includes two
position angles but the pointing angle includes only one\footnote{The
  pointing angle does have an additional source of statistical error
  due to uncertainty in the cluster centroid positions, which affects
  determination of the line connecting the two clusters.  However,
  this form of uncertainty is far less important than the position
  angle uncertainty when the clusters are separated by $R$ larger than
  a few cluster virial radii.}, and so comparatively the errors on the
pointing angle alignment are lowered by an initial factor of
$\sqrt{2}$.  Second, for each cluster pair, there are two pointing
angles but only one correlation angle.  As a result, each cluster pair
contributes twice to the pointing angle measurement, which reduces the errors on
it by another factor of $\sqrt{2}$.

\section{Comparison with theory}\label{S:theory}

We compare these observations with the theoretical predictions
for a $\Lambda$CDM cosmology from \cite{2005ApJ...618....1H}.  That study utilised a $2 \times
10^9$ particle $N$-body simulation to study the evolution of cluster
ellipticity, orientation, and alignment for $0<z<3$. The simulation
box length was $1500h^{-1}$Mpc, resulting in a large
simulated cluster sample.  The force softening length was
$17h^{-1}$kpc, so the spatial resolution was more than sufficient to
resolve the cluster halos. Only masses above $2 \times
10^{13}h^{-1}M_{\sun}$ were considered as clusters.

The observations include many systematic errors that are not 
in the simulation.  The most important ones are (1) the line-of-sight
redshift selection and complications due to photometric redshift
error, (2) cluster centroiding error (misidentification of the BCG
due to some 
algorithmic error), (3) noise in the determination of the cluster
position angle due the small number of cluster member 
galaxies ($\ge 5$) used to estimate the cluster shape, and (4)
  the need to exclude clusters that appeared nearly round, due to the
  difficulty in estimating a position angle. The first three of these systematic
errors will reduce our ability to detect intrinsic alignments and
weaken the observed signal.  Thus,
we do not compare with the direct predictions of 
\cite{2005ApJ...618....1H} (for their $0<z<0.5$ sample), but rather
with reduced predictions as described below.

The line-of-sight pair selection criterion, $|\Delta
  z_\mathrm{phot}| < 0.015$, was chosen to balance competing
  considerations.  In the absence of photometric redshift error, we
  would ideally attempt to mimic the \cite{2005ApJ...618....1H}
  selection of pairs within $100h^{-1}$Mpc; even with spectroscopic
  redshifts this would be complicated by redshift-space distortions, but
  we could at least hope to come fairly close to what was done in the
  simulations.  However, the photometric redshift errors correspond to
  typical separations of $\sim 50$\hmpc, making it impossible to
  imitate a strict line-of-sight separation.  We could simply choose a
  $\Delta z$ corresponding roughly to $100$\hmpc, but empirical tests
  showed that the contamination from completely unassociated clusters
  along the line-of-sight became unacceptably large.  These
  unassociated clusters dilute the expected signal and therefore the
  detection significance since their orientations are purely random,
  and indeed, even with our chosen $\Delta z_\mathrm{phot}$, we must
  impose a correction for it (to be described below).  We therefore
  err on the conservative side and use a $\Delta z_\mathrm{phot}$
  corresponding to roughly the $1\sigma$ photometric redshift error in
  order to be able to measure the signal.

The contamination due to accidental inclusion of unassociated clusters
because of photometric redshift error can be
estimated via simulation.  We simulate galaxy clusters with constant comoving number
density, assign a photo-$z$ assuming $\sigma(z_\mathrm{phot}) =
0.015$ (Gaussian), and estimate what fraction of the clusters within
$|\Delta z_\mathrm{phot}| = 0.015$ are actually $>100h^{-1}$Mpc
apart on the line-of-sight (with this separation chosen because it is
the criterion used by \citealt{2005ApJ...618....1H}).   Given a
contamination fraction $0<\Gamma<1$ defined as the fraction of
  cluster pair candidates that satisfy our $|\Delta z_\mathrm{phot}|$
  cut but that are more than $100$\hmpc\ apart along the
  line-of-sight, 
and theoretical predictions $\langle
\cos^2\theta_{c,p}\rangle_{\Lambda\mathrm{CDM}}$, we compare our measured signals with
\begin{equation}
\langle\cos^2 \theta_{c,p}\rangle_\mathrm{smeared} = 0.5\Gamma+ (1-\Gamma)\langle\cos^2\theta_{c,p}\rangle_{\Lambda\mathrm{CDM}}.
\end{equation}
We cannot estimate $\Gamma$ from the random, simulated clusters alone,
because that simulation only tells us the relative contamination when
the intrinsic (3d) cluster correlation function $\xi\ll 1$.  When clustering is
significant, the relative contamination becomes smaller.  We quantify
this effect by using the simulated contamination fraction $\beta=0.35$
in the absence of clustering, 
and the observed (projected) cluster correlation function.  The
simulated contamination fraction can be defined $\beta = N_S/(N_S+N_R)$
where $N_S$ represents spurious cluster pairs that appear within our
$\Delta z_\mathrm{phot}$ due only to photometric
redshift error, and $N_R$ represents those real pairs that are expected due to a purely
random distribution of clusters in the survey volume.  What we really
care about is $\Gamma = N_S/(N_S+N_R+N_E)$ where $N_E$ represents the excess
cluster pairs that are there in reality due to a non-zero cluster correlation
function.  Fortunately, we also measure the projected correlation
function, 
\beqa
w(R) +1 &=& \frac{N_\mathrm{pairs}\text{ in real
    data}}{N_\mathrm{pairs}\text{ in random cluster
    catalogue}} \notag\\
 &=& \frac{N_R+N_S+N_E}{N_R+N_S}.
\eeqa
To estimate $\Gamma$ we then use
\beq
\Gamma(R) = \frac{\beta}{1+w(R)}.
\eeq
In the case that clustering is insignificant, as on large scales, $w\approx 0$ and
$\Gamma(R)=\beta=0.35$. On the smallest scales, we find $w\approx 2 $ and
therefore $\Gamma\approx 0.12$. This
calculation 
assumes a Gaussian photo-$z$ error distribution, which is unlikely to
be valid in detail.  Large tails in the photo-$z$ error distribution
would tend to further weaken the observed signal.  

The theoretical predictions shown in Figs.~\ref{fig:corr_results} and~\ref{fig:point_results} have been
multiplied by this scale-dependent correction factor for photometric redshift error.
Still, the observed signals (especially the pointing angle alignment) are weaker 
than the theoretical predictions, even accounting for this
contamination.  The correlation alignment is marginally
consistent with the theoretical predictions.  For
the pointing angle, while the alignments are strongly detected
($\sim 10\sigma$), they are considerably weaker than  the 
theoretical predictions. 
	
The effect of the second observational error considered here, cluster
centroiding error, is more difficult to model accurately.  While
maxBCG centroiding errors have been modeled on mock catalogues
\citep{2007arXiv0709.1159J}, realistic cluster centroiding errors have
only been estimated for special cluster subsamples such as
the very massive ones that have strong X-ray detections (e.g., \citealt{2009ApJ...697.1358H}).  Thus, modeling this effect is
beyond the scope of this paper, and we merely state that
it should reduce the observed correlations.

Third, we estimate the impact of computing the cluster
  ellipticity and position angle from only $\ge 5$ (typically 5--6)
  galaxies.  In principle, this introduces measurement error in the
  position angles that is typically 15 degrees
  \citep{2010MNRAS.405.2023N}, which will dilute the predicted signal
  (it also increases the noise, but we have correctly accounted
  for this in our estimation of errorbars already).    
While the effect of statistical error in the cluster position angles
is in principle complicated, we can appeal to simple arguments to
roughly estimate its impact.  It should result in the true
distribution of correlation or pointing angles, $p(\cos^2{\theta_{c,p}})$,
being convolved with some error distribution, which we assume to be
Gaussian.  Unfortunately, \cite{2005ApJ...618....1H} do not report a
full distribution of $\cos^2{\theta_{c,p}}$, only its mean value.  Thus, to estimate the effect on our statistic
$\langle\cos^2{\theta_{c,p}}\rangle$, we 
  arbitrarily assume that $p(\cos^2{\theta_{c,p}}) = A + B
  \cos^2{\theta_{c,p}}$.  We fix the values of $A$ and $B$ by imposing
  two requirements: 
  that the
  probability distribution be normalised to $1$, and that 
  $\langle\cos^2{\theta_{c,p}}\rangle$ should be consistent with the simulations.  Then, we convolve
  that distribution with a Gaussian distribution with
  $\sigma_{\mathrm{obs},c,p}$ (where $\sigma_{\mathrm{obs},c}
  = \sqrt{2} \sigma_{\mathrm{obs},p}$ since the former includes two
  position angles and the latter includes one).  In this simple limit,
  for small $\sigma_{\mathrm{obs},c,p}$, 
  we expect an observed correlation
\beq
\langle\cos^2\theta_{c,p}\rangle_\mathrm{obs} = 0.5 +
e^{-2\sigma_{\mathrm{obs},c,p}^2} \left[\langle\cos^2{\theta_{c,p}}\rangle- 0.5\right].
\eeq
This reduces to
$\langle\cos^2\theta\rangle_\mathrm{obs}=\langle\cos^2{\theta}\rangle$
as $\sigma_{\mathrm{obs},c,p}$ approaches zero.  Following
\cite{2010MNRAS.405.2023N}, if we assume that the position
angle errors are typically $\sim 15$ deg, then the correlations
are reduced by $\sim 15$ per cent.  This reduction cannot account for the apparent difference between theory and
observation in Fig.~\ref{fig:point_results}, which means that
reconciling this detection with the theory will require more detailed
modeling of observational systematic errors.

Finally, we consider the impact of excluding those clusters
  that are within $1\sigma$ of being round and therefore have poorly
  defined position angles.  There are two options to consider.  The
  first is that those clusters truly are round (typically cluster dark
  matter halos are triaxial in $N$-body simulations, but could appear
  nearly round due to projection at certain position angles).  In that
  case, the theoretical predictions for the pointing and correlation
  angle statistics include a contribution of zero from such round
  clusters, and our exclusion of them will artificially inflate the
  signal.  The second option is that the clusters are not round, but
  that they appear so due to noise given that typically only 5--6
  galaxies are used to define the shape; this option likely represents
  the majority of the clusters that were excluded.  In that case, the
  theoretical predictions include some real alignment signal for these
  clusters, which cannot be measured in reality since their position
  angles are too noisy.  If we were to include them, it would dilute
  the measured signal.  Excluding them is the proper choice in this
  case, and as long as those clusters that are excluded are a
  representative subsample with respect to intrinsic alignment
  properties, then it should not cause any bias in the signal. 
  Nonetheless, future work should include a more careful simulation
  with such effects directly incorporated before generating
  theoretical predictions.
 
\section{Discussion}

In this paper, we have presented a strong detection of galaxy cluster
intrinsic alignments to very large scales of $100h^{-1}$Mpc,
representing a tendency of clusters to
point preferentially towards other clusters.  Depending on the method
used to select clusters and assign photometric redshifts, the strength of the
detection (averaged over $1<R<100h^{-1}$Mpc) ranges from $6\sigma$ to
$10\sigma$.  The alignment of pairs of cluster position angles with
each other was only detected at the
$2$--$2.5\sigma$ level.  This measurement constitutes the first strong 
detection of the intrinsic alignments of clusters to $100h^{-1}$Mpc
with a large, statistical cluster sample ($6$--$8\times 10^3$ clusters) using a uniform set
of photometric data.

The observed correlation angle alignment is consistent, within the
large error bars, with the theoretical $\Lambda$CDM prediction.  The
pointing angle alignment, while strongly detected to $100h^{-1}$Mpc,
is weaker than expected; this is likely due to various systematic
observational uncertainties, all of which tend to weaken the observed
signal.  We find that photometric
redshift error cannot fully account for the dilution of the signal.
Future work may include an exploration of how the signal varies with
the BCG dominance, to test whether merging clusters might be
responsible for diluting the alignment signal.  It would also be
helpful to carry out this measurement using a sample spanning a
broader redshift range, to check for redshift evolution.  These future
investigations can be used to understand the primary source of
the alignments, i.e. whether they are primordial or due to tidal
torques from the large-scale density field that would lead to an increase at later times.

A
more detailed comparison of these observations with $\Lambda$CDM would require simulations
that fully incorporate cluster selection effects (including false
cluster identification, incorrect division of larger clusters into two
smaller clusters, etc.), centroiding errors, the shape estimation
procedure including the elimination of round clusters, and position angle
uncertainties.  This level of detail is beyond the scope of this work.
However, we have demonstrated the power of large imaging surveys such
as SDSS to detect for the first time the cluster intrinsic alignments
to very large scales of $100h^{-1}$Mpc, which can test the $\Lambda$CDM prediction of massive dark matter halo
alignments.  This finding suggests that large imaging surveys planned
for the near future, and eventually the Large Synoptic Survey
Telescope (LSST), should be successful in mapping out 
cluster intrinsic alignments as a function of cluster mass and redshift.

\section*{Acknowledgments}

The authors would like to thank the anonymous referee for constructive
feedback that led to improvements in the quality of this work.    
We thank Michael Strauss for providing useful comments about this project.

\bibliography{cosmo,cosmo_preprints,lensing,ms_RR,newrefs,sdss,tullyfisher}

\end{document}